# Optimizing Floating Locations in Hard Disk Drive by Solving Max-min Optimization

Victor Liu[1*], Hongtao Yang[2], Haiming Li[1], Chifu Yang[3]
[1]Department of Mechanical Engineering, Beijing Jiaotong University, Beijing, 100044, China
[2]Department of Petroleum & Geosystems Engineering, University of Texas, Austin, TX, 78712, U.S.A
[3]Harbin Institute of Technology, Harbin, 150006, China

## Abstract

Floating operation is very critical in power management in hard disk drive (HDD), during which no control command is applied to the read/write head but a fixed current to counteract actuator flex bias. External disturbance induced drift of head may result in interference of head and bump on the disk during drifting, leading to consequent scratches and head degradation, which is a severe reliability concern in HDD. This paper proposes a unique systematic methodology to minimize the chances of hitting bump on the disk during drive floating. Essentially, it provides a heuristic solution to a class of max-min optimization problem which achieves desirable trade-off between optimality and computation complexity. Multivariable nonlinear optimization problem of this sort is reduced from *NP*-hard to an arithmetic problem. Also, worst-case is derived for arbitrary bump locations.

## 1. Introduction

In HDD, actuator moves head towards certain track on the rotating disk to perform read/write operation as shown in Fig. 1. While HDD is not performing any read/write operation, head is usually floating on the disk to save energy. Head floating is very critical in power management as consumers are more stringent in power consumption of HDD nowadays. During head floating, no control command is applied to the actuator but a fixed current to counteract actuator flex bias. Due to the external disturbance and system uncertainty, head may move uncontrollably and drift towards high thermal asperity (HTA), prominent bumps on the disk. HTA is inevitable on the disk during high volume manufacturing (HVM) of HDD. If head interferes with HTA in its path, the consequent scratches and head degradation may severely damage reliability of drive.

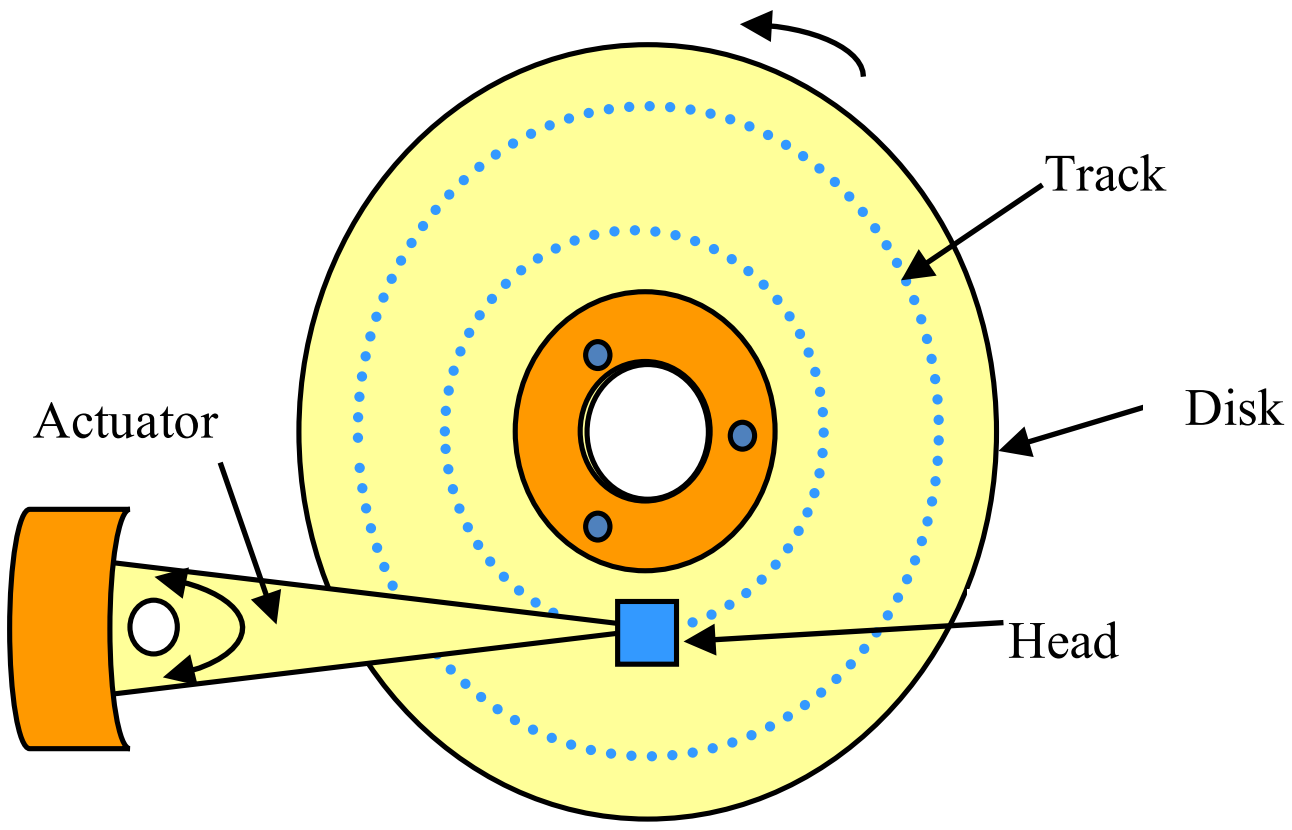

Figure 1: HDD scheme.

In HDD, head degradation and scratch in the media due to head disk interference (HDI) poses a great concern for drive reliability [1,2]. The prior-art practice is to select floating locations empirically for high volume drives [3,4]. However, fixed floating locations are not optimized for each drive due to variations of HTA locations by drives, increasing chances of HDI among high volume production. Therefore, this empirical practice would unavoidably hit yield during HVM. This paper addresses the aforementioned issues by proposing a systematic approach in designing head floating locations. The crucial factor is to optimize floating locations by maximizing minimum distance between head floating and HTA position. Essentially, this methodology provides a heuristic solution to a class of *NP*-hard nonlinear optimization problem, which achieves desirable trade-off between optimality and computation complexity.

## 2. Problem statement

To ensure drive reliability, floating locations should be placed far away from bumps (i.e. HTA), each other and boundaries. Essentially, we need to maximize following objective function,

$$\text{Objective function: } \max_{p_1,\dots,p_{NFLs}} (\min_{i,j,i\neq j}(|p_i - p_j|)) \tag{1}$$

$$\text{s.t. } B_l < p_i < B_u \text{ for each } i \in \{1,\dots,NFLs\}$$

where $p$ is position of point of interest (floating points, HTAs and boundaries); $i,j = 1,\dots,NFLs + NTAs + 2$, where $NTAs$ and $NFLs$ are number of HTA and floating locations, respectively. $i = 1,\dots,NFLs$ and $i = NFLs+1,\dots,NFLs+NTAs$ corresponds to $i$ th floating point and HTA, respectively and last 2 points correspond to boundaries. $B_l$ and $B_u$ are lower and upper bounds for boundaries, respectively. Given boundaries and bumps positions, floating locations need to be optimized from Eq. (1). The max-min optimization aims at constructing solutions to achieve the best possible performance in the worst-case scenarios, which is a typical nonlinear optimization problem [5]. Multivariable nonlinear programming problems are often difficult to solve [6-11]. Despite the rapid pace of algorithmic improvements, no generic algorithms can provide guarantees of success or of fast performance over a range of applications [23,24]. A variety of algorithms are proposed to solve a specific kind of complicated matters [12,13,25,26]. Numerical solution such as numerical search, e.g. genetic algorithm may take excessive time than needed [14]. It is noteworthy that execution time increases dramatically with more floating locations due to its *NP*-hard property.

In this paper, at first a heuristic solution is proposed to solve this class of max-min optimization problem. Multivariable nonlinear optimization problem of this sort is reduced from a *NP*-hard problem to an arithmetic problem. Statistically, the solution from proposed method can achieve quasi-optimality, as demonstrated in the later section. Also, worst-case is derived for arbitrary bump locations. At last, head waking-up frequency is derived based on the optimization of floating locations.

## 3. Proposed heuristic algorithm

Before the heuristic algorithm is presented, some notations are given for simplicity of following derivations. This max-min optimization task is regarded as allocations of floating locations into brackets consisting of boundaries and bumps. Area of each bracket refers to length between boundaries of bracket.

**Notations:**

$NFLs$: number of floating locations

$NTAs$: number of HTAs

$M$ : number of brackets

$Ls$ : total length of brackets, i.e. length of boundary

$m$: $m$ th bracket

$M$ : number of brackets

$A_m$: the area of $m$ th bracket

$N_m$: number of floating locations in $m$ th bracket

Based on above definitions, $M = NFLs + NTAs + 1$. Furthermore, the spacing function for each bracket can be defined as,

$$s_m(N_m) = \frac{A_m}{N_m + 1} \tag{2}$$

where $m$ is $m$ th bracket and $N_m$ is number of floating locations in $m$; $A_m$ is the area of $m$ th bracket.

We can obtain best initial allocations of floating locations into brackets by forcing Eq. (3) and Eq. (4) as follows,

$$\frac{A_i}{N_i + 1} = \frac{A_j}{N_j + 1} \quad i, j = 1,2,...,M; i \neq j \tag{3}$$

$$\sum_{i=1}^{M} N_i = NFLs \tag{4}$$

where $N_i$ is the initial allocations of floating locations for $i$ th bracket among total $M$ brackets and we have

$$N_i = (NFLs + M) \cdot \frac{A_i}{Ls} - 1 \qquad i = 1,2,...,M \tag{5}$$

**Flowchart of optimization of floating locations**

Fig. 2 shows the flowchart of optimization process. At first, best initial number of floating points for each bracket is calculated from Eq. (5). Then the $i$ th brackets with $N_i < 1$, where $i = 1,2,...,M$ are eliminated and remaining brackets form new bracket group. On top of initial assignment of floating points to each bracket, $N_m$ for $m$ th bracket is adjusted until Eq. (4) is satisfied. The adjustment rule is as follows,

$$\text{If } \sum_{i=1}^{M} N_i < NFLs, \quad N_{m^*} = N_{m^*} + 1 \qquad \text{where } m^* = \arg\max_{m\in\{1,\dots,M\}} (s_m(N_m + 1))$$
$$\text{If } \sum_{i=1}^{M} N_i > NFLs, \quad N_{m^*} = N_{m^*} - 1 \qquad \text{where } m^* = \arg\min_{m\in\{1,\dots,M\}} (s_m(N_m)) \tag{6}$$

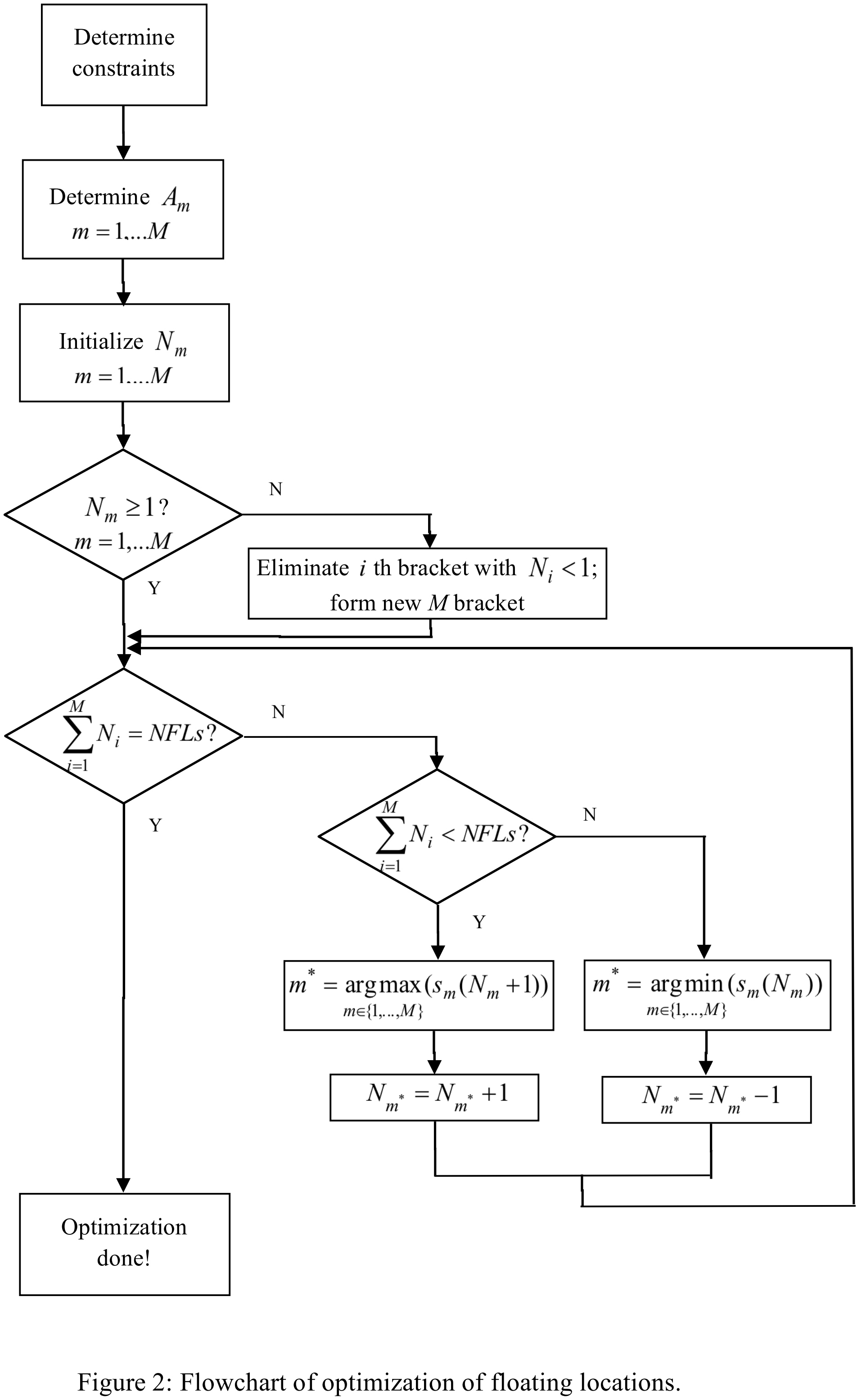

Figure 2: Flowchart of optimization of floating locations.

The proposed heuristic solution usually converges within several iterations of adjustment due to best initial assignment and it is insensitive to number of floating locations. That is, the proposed heuristic solution converts the originally *NP*-hard problem into a simple arithmetic problem. Reduction of this sort is similar to the conversion of stochastic control to deterministic control via inference from sensing measurement/observation [15-17] and conversion of infinite characteristic roots in time-delay system modeling into tractable stability map via root tendency analysis [18-22]. Related discussions regarding the analysis of the time-delay systems with infinitely many characteristic roots are available for the readers' reference [27-42]

**Worst-case spacing**

For given boundaries, number of floating locations and HTA, there are certain distributions of HTA within boundaries which lead to minimum objective function in Eq. (1). We need to figure out this worst-case spacing for minimum possible distance among floating locations, HTAs and boundaries. That is, for all possible HTA distributions within boundaries, we need to derive minimum of Eq. (1).

It can be derived formulated as that,

$$\min_{p_{NFLs+1},\dots,p_{NFLs+NTAs}} \left( \max_{p_1,\dots,p_{NFLs}} \left( \min_{i,j,i\neq j} \left( \left| p_i - p_j \right| \right) \right) \right) = \frac{Ls}{NFLs + 2M - 1}. \tag{7}$$

If floating positions cannot be optimized for each individual HDD, safe actuator drifting distance should be derived based on the worst-case scenario from HTA distribution from Eq. (7).

## 4. Head waking-up frequency

Drive usually relocates the actuator every fixed interval to ensure reliability during head floating. Based on minimum distance of floating locations, bumps and boundaries, minimum allowable head waking-up frequency can be determined by estimation of head drifting. head drifting can be modeled as follows,

$$m\ddot{x} + c\dot{x} = F \tag{8}$$

where $x$ is drifting distance, $m$ is mass of actuator-head assembly and $c$ is viscous coefficient. Force $F$ can be expressed as

$$F = f_b(x) - f_t(x) \cdot I + d \tag{9}$$

where $f_b$ and $f_t$ are flex bias and torque constant in terms of $x$, respectively. $d$ is external disturbance and $I$ is constant VCM current based on the flex bias learning on floating locations. Furthermore force $F$ can be reduced to for simplicity,

$$F = (K_b - K_t I) \cdot x + d \tag{10}$$

and head drifting can be modeled simply as mass-spring-damper system as,

$$m\ddot{x} + c\dot{x} + (K_t I - K_b) \cdot x = d. \tag{11}$$

Drifting distance of head with respect to time can be derived by solving second order differential equation from Eq. (11).
The related parameters in Eq. (11) can be roughly determined by experiments. Experimentally, for relocation period of $T_s$ = 60s, drifting distance can be up to ~2k tracks. We need to make sure following condition to avoid HDI between floating head and HTA,

$$\text{Distance between floating locations and HTA} > x(T_s) \tag{12}$$

where $x(T_s)$ is drifting distance of floating head from 0 to $T_s$. To achieve the optimal trade-off between power management and reliability, minimum allowable head-waking-up frequency can be determined by max-min the distance between floating locations and bumps.

## 5. Numerical results

### Case 1: 2 HTA and 1 seam track

In this case, bump locations consist of 2 HTA and 1 seam track, totaling 3 bumps between boundaries from 0 to 1k tracks. Six floating locations are required to be placed within the boundaries with largest minimum distance among floating locations, boundary and bumps. Fig. 3(a) shows the solution by proposed heuristic method with minimum distance of 98. Fig. 3(b) shows CDF of randomly generated floating locations. From Fig. 3(b), distance of 98 is greater 99.8% of randomly generated solutions, demonstrating quasi-optimal solution by proposed method.

### Case 2: 3 HTA and 1 seam track

In this case, bump locations consist of 3 HTA and 1 seam track, totaling 4 bumps between boundaries from 0 to 1k tracks. Six floating locations are required to be placed within the boundaries with largest minimum distance among floating locations and bumps. Fig. 4(a) shows the solution by proposed heuristic method with minimum distance of 84. Fig. 4(b) shows CDF of randomly generated floating locations. From Fig.

4(b), distance of 84 is greater 99.7% of randomly generated solutions, demonstrating quasi-optimal solution by proposed method.

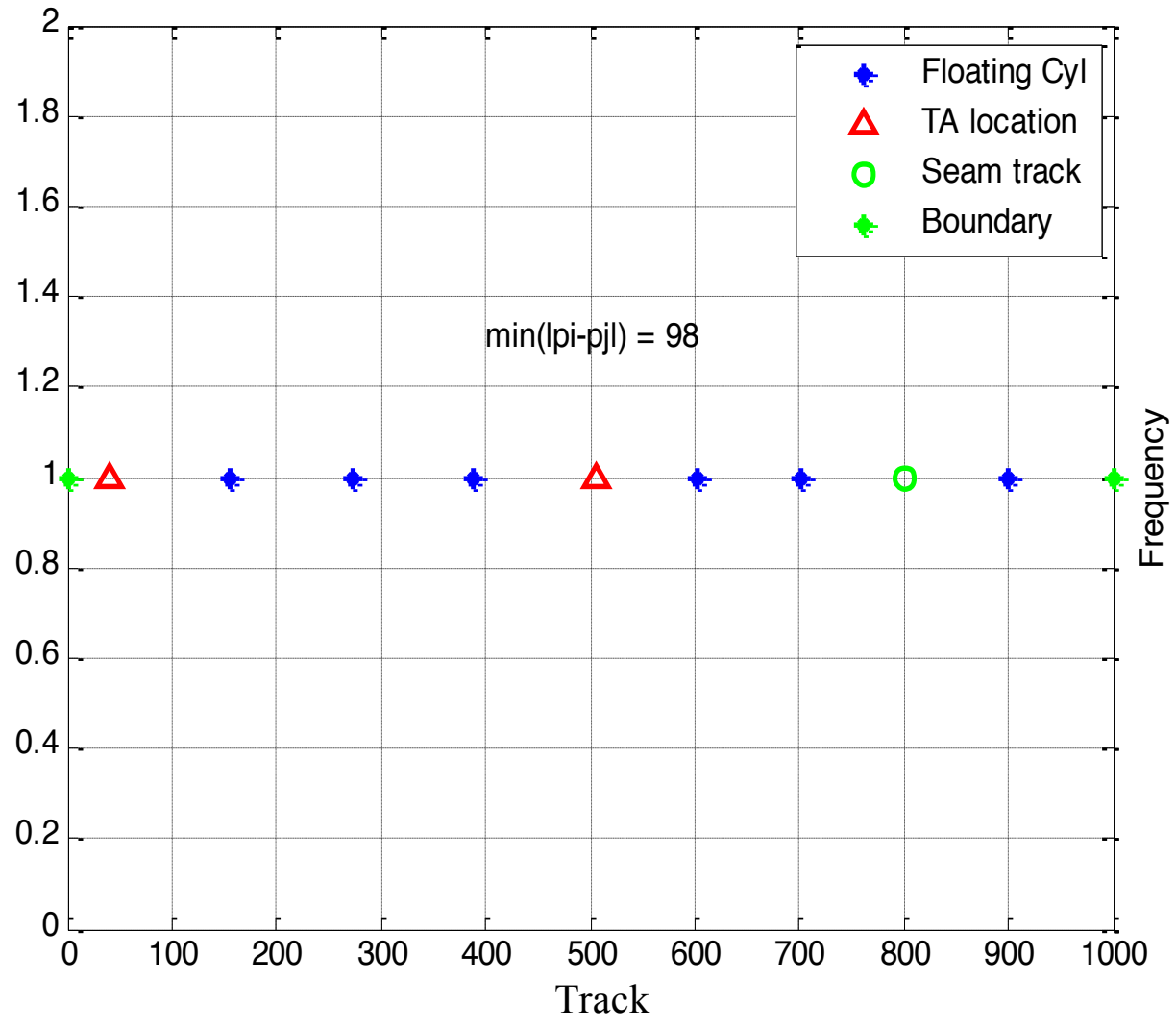

Figure 3(a): Floating locations profile for case 1.

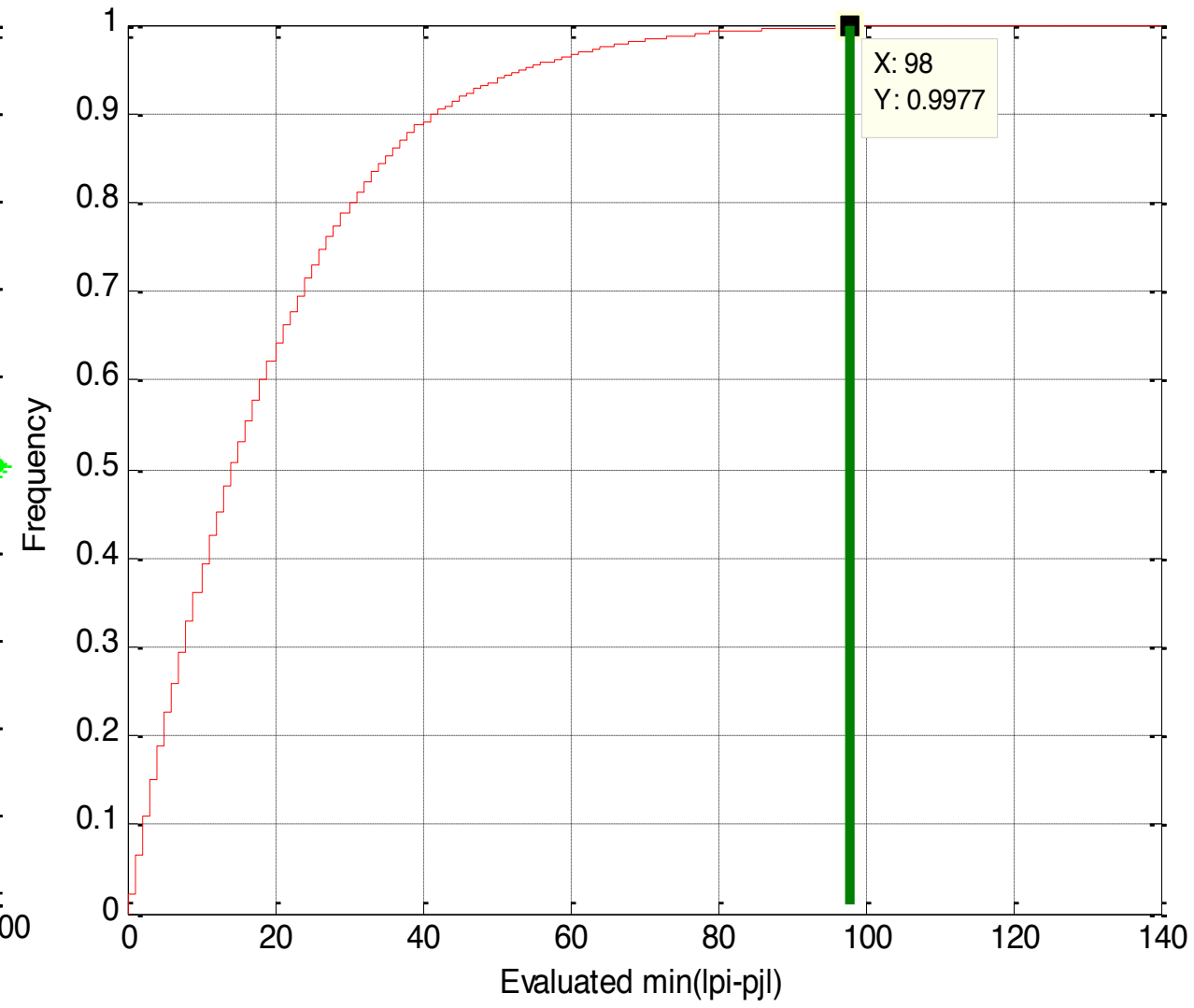

Figure 3(b): CDF of randomly generated floating positions for case 1.

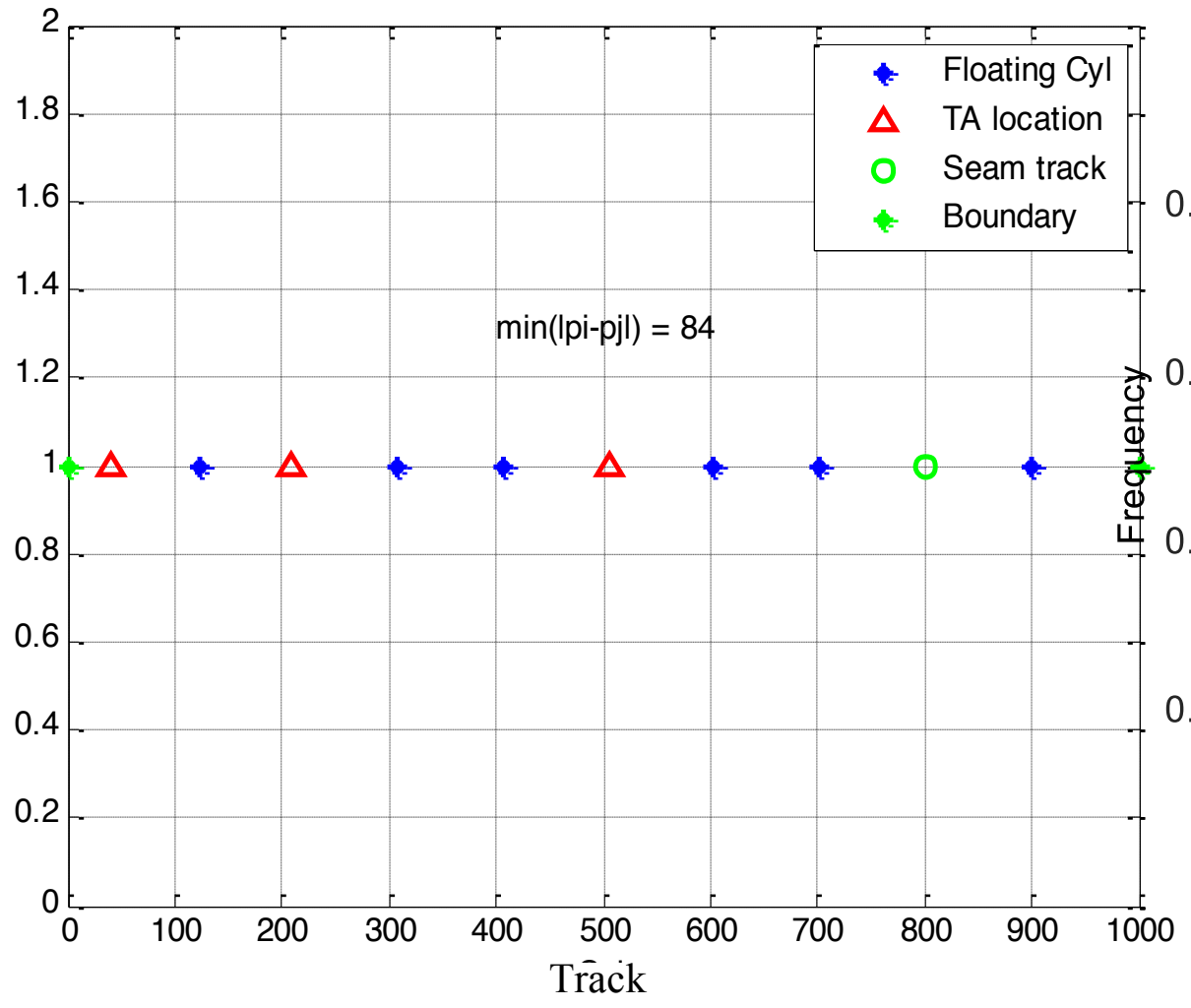

Figure 4(a): Floating locations profile for case 2.

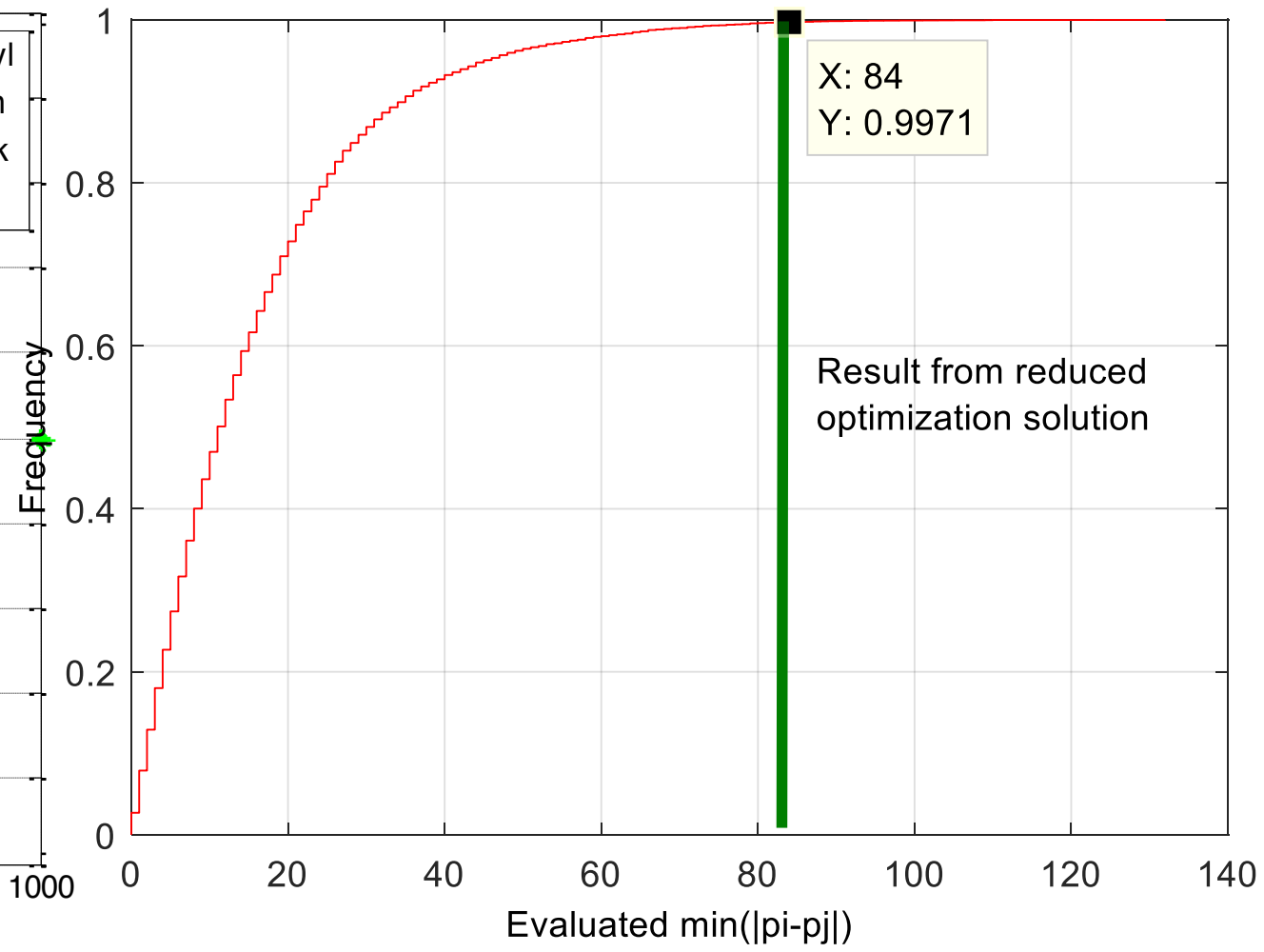

Figure 4(b): CDF of randomly generated floating locations for case 2.

## 5. Conclusion

Floating operation is very critical to power saving in hard disk drive (HDD), during which no control command is applied to the actuator but a fixed current to counteract actuator flex bias. External disturbance induced drift of head may result in interference of head and bump on the disk during drifting, leading to consequent scratches and head degradation, a severe reliability concern in HDD. This paper proposes a unique heuristic methodology to minimize the chances of hitting bump on the disk during drive floating.

Essentially, given bump locations it provides a solution to a class of max-min optimization problem. Also, worst-case is derived for arbitrary bump locations. Validations demonstrates the heuristic solution from proposed method can achieves desirable trade-off between optimality and execution time.